%% file: p.tex
\documentclass[letterpaper,twocolumn,10pt]{article}

\usepackage{usenix}
\usepackage[square,comma,numbers,sort&compress]{natbib}

\include{pkgs}

\newcommand{\sys}{\mbox{\textsc{DriveShaft}}\xspace}

\newenvironment{tightitemize}
 {\begin{list}{$\bullet$}{
                \setlength{\leftmargin}{10pt}
		\setlength{\itemsep}{0pt}
		\setlength{\parsep}{0pt}
		\setlength{\topsep}{0pt}
		\setlength{\parskip}{0pt}
		}
 }
 {\end{list}}

\newcounter{tecounter}
\newenvironment{tightenumerate}
 {\begin{list}{\arabic{tecounter}.}{
                \usecounter{tecounter}
                \setlength{\leftmargin}{10pt}
		\setlength{\itemsep}{0pt}
		\setlength{\parsep}{0pt}
		\setlength{\topsep}{0pt}
		\setlength{\parskip}{0pt}
		}
 }
 {\end{list}}

\input{cmds}

\input{rev}

\begin{document}

\input{hdr}

\date{}
\maketitle

\input{abstract}
\input{intro}

\input{overview}
\input{design}
\input{impl}

\input{eval}

\input{relwk}
\input{conclusion}

\balance
\bibliographystyle{abbrvnat}
\footnotesize
\setlength{\bibsep}{3pt}
\bibliography{driveshaft}
\end{document}

%% file: pkgs.tex
\input{warning}

\usepackage[hyphens]{url}
\usepackage[breaklinks,colorlinks]{hyperref}
\usepackage[usenames,dvipsnames]{xcolor}
\hypersetup{citecolor=blue,linkcolor=blue}
\usepackage{amsmath,amsopn,amssymb}
\usepackage{subfigure}
\usepackage{endnotes,microtype,xspace,graphicx,fancyvrb,multirow}
\usepackage{booktabs}
\usepackage{array,underscore,relsize}
\usepackage[T1]{fontenc}
\usepackage{times}
\usepackage{fancyhdr,lastpage}
\usepackage{enumitem}
\usepackage[labelfont=bf,font=small,skip=5pt]{caption}
\usepackage{fp}
\usepackage{siunitx}

\usepackage{balance}

%% file: warning.tex
\usepackage{silence}

%% file: cmds.tex
\fvset{fontsize=\scriptsize,xleftmargin=8pt,numbers=left,numbersep=5pt}

\input{code/fmt}

\def\Snospace~{\S{}}

\newif\ifdraft\drafttrue
\newif\ifnotes\notestrue
\ifdraft\else\notesfalse\fi

\input{glyphtounicode}
\newcolumntype{R}[1]{>{\raggedleft\let\newline\\\arraybackslash\hspace{0pt}}p{#1}}

\newcommand{\squishlist}{
\begin{itemize}[noitemsep,nolistsep]
  \setlength{\itemsep}{-0pt}
}
\newcommand{\squishend}{
  \end{itemize}
}

\usepackage{tikz}

\usepackage{xstring}
\newcommand{\PP}[1]{
\vspace{2px}
\noindent{\bf \IfEndWith{#1}{.}{#1}{#1.}}
}



%% file: rev.tex
\gdef\therev{}
\gdef\thedate{}

%% file: hdr.tex
\title{\sys: Improving Perceived Mobile Web Performance}

\ifdefined\DRAFT
 \pagestyle{fancyplain}
 \lhead{Rev.~\therev}
 \rhead{\thedate}
 \cfoot{\thepage\ of \pageref{LastPage}}
\fi


\author{
{\rm Ketan Bhardwaj, Ada Gavrilovska}\\
	Georgia Institute of Technology
\and
{\rm Moritz Steiner, Martin Flack, Stephen Ludin}\\
	Akamai Technologies
} 

%% file: abstract.tex
\begin{abstract}
With mobiles overtaking desktops as the primary vehicle of Internet
consumption, mobile web performance has become a crucial factor for
websites as it directly impacts their revenue. In principle, improving
web performance entails squeezing out every millisecond in the webpage
delivery, loading and rendering. However, on a practical note, an
illusion of faster websites suffices. This paper presents
'DriveShaft', a system envisioned to be deployed in Content Delivery
Networks, which improves the perceived web performance on mobile devices
by reducing the time taken to show visually complete webpages, without
requiring any changes in websites, browsers, or any actions from end
user. DriveShaft employs (i) crowd-sourcing, (ii) on-the-fly
JavaScript injection, (iii) privacy preserving desensitization, and
(iv) automatic HTML generation to achieve its goals. Experimental
evaluations using 200 representative websites on different networks
(Wi-Fi and 4G), different devices (high-end and low-end phones) and
different browsers, show a reduction of 5x in the time required to see
a visually complete website while giving a perception of 5x-6x faster
page loading.
\end{abstract}

%% file: intro.tex
\section{Introduction}
\label{sec:intro}

Faster web response is a critical factor in any Internet-based business. It has direct impact on user experience, revenue, costs, and discover-ability (page rank). Modern websites try to engage users with a fastest possible visual response to avoid loosing users attention. For mobile, it is even more important due to fragmented and shorter attentions user span in use of mobile device vs. desktop~\cite{shorterattentionspan} and is more difficult to achieve because of inherent resource constraints (compute, battery, data costs etc.) in mobile devices. This has led to a conservative approach being adopted to resource usage~\cite{bestprac} in mobile web browsing.

Practical lower bounds of web performance are governed by a large number of factors in addition to the physical limitations of wireless links. Those factors include (i) latency of $\sim$600 ms\footnote{DNS lookup (1 x RTT) + Connection set-up (2 x RTT) + Request-Response (1 x RTT) = 4 x RTT  = 600 ms (Assuming RTT = 150 ms including the mobile back-haul latency)} comprising of DNS lookup time, connection set-up time and first request-response cycle; (ii) increasing complexity and/or size of richer websites and (iii) latency introduced by use of solution like data compression proxies~\cite{flywheel} etc. This leaves only 1400 ms at disposal to fetch web page components (HTML, images, java scripts etc.), parse HTML, execute Java-scripts, render the web page etc., to show a visually complete page to users on mobile devices. On a more practical note, to keep users engaged an illusion of faster visual response is often considered sufficient as opposed to actually web page rendering.

Extensive previous work in the field of web performance has focused on optimizing different steps involved in web page loading by reducing DNS lookup time~\cite{dnsopt1, dnsopt2}, re-using persistent connections, web object pre-fetching~\cite{prefetching1,prefetching2}, optimizing JavaScript execution 
\cite{jsopt1,jsopt2}, rendering~\cite{grigorik2013high}, etc. More recent work has developed techniques to manage inter-dependencies in loading a web page to reduce idle time~\cite{wprof, othernsdirefs}  and to selectively offload web page processing~\cite{offloading} to improve mobile web performance. At the networking protocol layer, new experimental protocols have been introduced, such as SPDY, HTTP 2.0, QUIC, etc., to address web performance concerns. State of art research has proposed the use of data compression proxies~\cite{flywheel} to reduce the cost of mobile web browsing by reducing the number of costly bytes transferred to end user device. In wake of extensive prior work, the rationale for DriveShaft (DS) is derived from the following observations (\textsection\ref{sec:moti}):

\begin{tightitemize}
\item With growing structural complexity of web pages, achieving faster perceived web performance seems an impossible goal. Proposed approaches take a micro-optimization approach. 
They focus on optimizing individual steps in the web page loading, such as intelligently resolving dependencies, pipelining network and compute bound activities, rendering order, etc.. Since rendering a single web page entails fetching a large number of elements from different domains, resolving their inter-twined dependencies and rendering them on a device screen, 
the complexity of these steps limits the effectivness of such micro-optimization approaches. 
In this paper, we ask a different question, i.e., "Can we decouple the structural complexity of web pages from its load time?"

\item To make the Internet seem instant, we have to remove the page load latency completely. This is fundamentally impossible as 
there is inherent latency in network access and in performing computation on a mobile device. Instead, we ask the question, "How can we hide this latency from the end users?''. This is in contrast to most existing solutions that try to minimize, but continue to expose this latency in the critical path of the user-perceived web site performance. 

\item Further, we make a seemingly obvious but important observation that there exists orders of magnitude difference in the frequency of visual changes in popular web sites vs. the frequency of web site accesses. This provides a time window to potentially harvest web page rendered on one user device and use it to provide faster response to other users. 
\end{tightitemize}

In this work, our goal was to create a system that:
\begin{tightitemize}
\item safely crowd sources and harvests web page screen shots from end client devices;

\item using harvested snapshots, 
makes it possible to decouple the {\em time to show a visually complete web page} from the web page's structural complexity; and  

\item does not require changes in the way websites operate today.
\end{tightitemize}

The answers to the above questions along with design exploration for making them possible constitutes the technical contribution of this paper. Specifically, this paper describes \sys (DS) -- a system for faster delivery of visually complete web pages. We envision its deployment in Content Delivery Networks (CDNs). 
%
\sys represents an approach where users are shown a interstitial visual pop of a screenshot of the requested web page on their device. The snapshot is as close as possible to a pixel perfect copy of the fully rendered version of that web page. When the screenshot is being displayed, the request for the actual web page is the triggered in the background (i.e., in a tab invisible to end user) which is then rendered on a invisible tab and is flipped with the snapshot on rendering completion. Websites using \sys can allow end users to opt-in if they are willing to post to \sys servers the snapshot of their browser while visiting a website, which can later be used to service other users. This however, neither requires any change in browsers in end user devices nor change in  required cange existing website need not be changed and/or be aware of \sys. 
Techniques that make \sys possible include: 

\begin{tightitemize}
\item crowd sourcing to harvest web page snapshots.
\item on-the-fly JavaScript injection in web pages at CDN nodes.
\item privacy preserving desensitization of harvested snapshots.
\item automatic generation of HTML pages comprised of a click map containing links same as the original web page, overlaid on its snapshot, where 
\item the HTML page also contains a pre-render hint to the original web page, along with logic to redirect to original web page when pre-rendering is completed(~\autoref{sec:arch}).
\end{tightitemize}

The experimental evaluations in \textsection\ref{sec:eval}, using 200 websites from Alexa's list show that with \sys, web sites can achieve up to 5x improvement in their speed index, a measure for how quickly a web page is complete visually~\cite{speedindex}, and reduced perceived web page load time by 5x to 6x. Further, the results also suggest that those improvements are visible (i) when using cellular (4G) or Wi-Fi, (ii) more prominent in low-end vs. high-end devices, and (iii) persist with different browsers. We compare \sys's performance with the best case of loading a web page from the device's local web cache and Flywheel Google's data compression proxy to show that if used together, 
\sys can hide the increase in time to first paint while reducing data burdens on end users. 
\sys is able to achieve these benefits without any changes in websites by web developers or any actions from end user, and with modest overheads on client device. We discuss technology-related concerns with the \sys approach in \textsection\ref{sec:disc}. Related work appears in \textsection{\ref{sec:related}} with the concluding remarks in \textsection{\ref{sec:conc}}.

%% file: overview.tex
\section{Motivation}
\label{sec:moti}
\begin{figure}[t]
\centering
\includegraphics[trim=30mm 40mm 15mm 25mm, width=0.9\columnwidth]{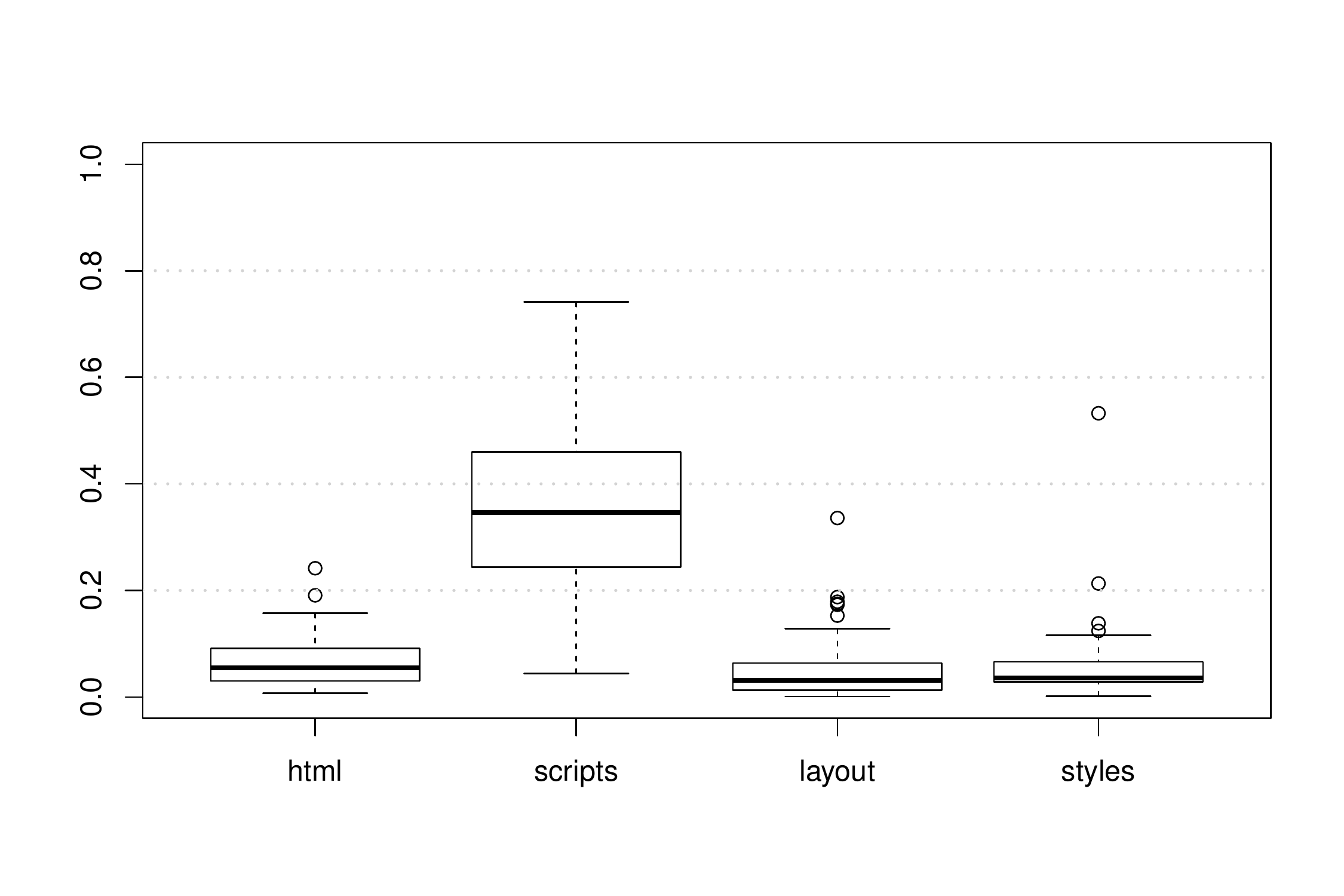}
\vspace{4ex}
\caption{Computationally intensive website rendering: Fraction of the overall page load time spent on CPU intensive tasks observed in Alexa's top 500 website}
\label{fig:comp}
\end{figure}
The motivation behind the design of \sys system is extracted from the following goals and observations about the current state of operation of the web:

\noindent {\bf Reducing time to show visually complete web page:}
Web experience especially on mobile devices relies more on the perception of how fast a web page is rendered rather than the actual web page rendering time. Examples of current state of art of techniques deployed to improve perceived performance include but are limited to (i) ensuring there is some response to users, e.g., spinning wheel to engage users, (ii) in-lining css in HTML required to render web page, (iii) fetching only the components required to render a web page up to a single view fold before starting rendering, etc. Simply put, if an end user is provided with a perception of faster web page rendering, it has similar affect on web experience as improving web performance. From this, we derive our motivation behind the potential use of a DS proxy to {\em reduce web page's perceived performance or time to show visually complete web page} on mobile devices.

\noindent{\bf Decoupling web page's time to show visually complete page from its structural complexity:}
Data from HTTP archive suggests a continuous increasing trend in (i) website total transfer sizes, (ii) number of requests per web page, (iii) an increase of JavaScript transfer size, often paralleled with increase in compute required for web page rendering, and (iv) a increase in number of Javascripts per page. Note that in both cases, if the response is compressed e.g., by using a data compression proxy, the transfer size is smaller than the original uncompressed content shown but at the cost of additional computation on mobile device to decompress the content. This is result of push to make web interactions richer for end clients resulting in increased inter-dependencies~\cite{wprof} for a single web page. In any case, we can safely assume that the websites are becoming more and more complex. We posit that this would put pressure on web loading mechanisms especially on mobile devices with limited compute capabilities. From this, we derive our motivation behind designing \sys in way that makes it possible to {\em decouple the structural complexity of a web page from its time to show visually complete web page}.

\noindent{\bf Hiding network and computational overhead in web performance:}
Previous research on web performance suggests that web page load time and hence, time to show visually complete web page is adversely affected by the amount of computation~\cite{offloading,wprof} required for rendering the web page in addition to well known network delays. Specifically, on-device computation seems to be a bottleneck in web page load time for more than 30\% of the popular websites\cite{wprof} and it is exacerbated when using a slow machines (mobile devices) vs. fast machines (desktop grade machines) where the fast and slow machines differ in speeds by a factor of 8x~\cite{resusage}. To verify these observations, we measured the fractions of overall page load time which are spent in CPU intensive tasks, we can see from Figure~\ref{fig:comp} that for the websites that spend more time in computation, compute is a major factor in their time to first paint. Also, important to note here is that data compression proxies e.g., Google Flywheel~\cite{flywheel} increase computational overhead on mobile devices which ends up degrading web performance on mobile devices in terms of increase in latency which is prominent for smaller web pages. From this, we derive our motivation of designing \sys as a system that {\em hides the time required in network and/or computation in web page rendering while keeping users engaged in meaningful ways}. 

\begin{figure}[t]
\centering
\includegraphics[trim=10mm 10mm 15mm 25mm, width=0.9\columnwidth]{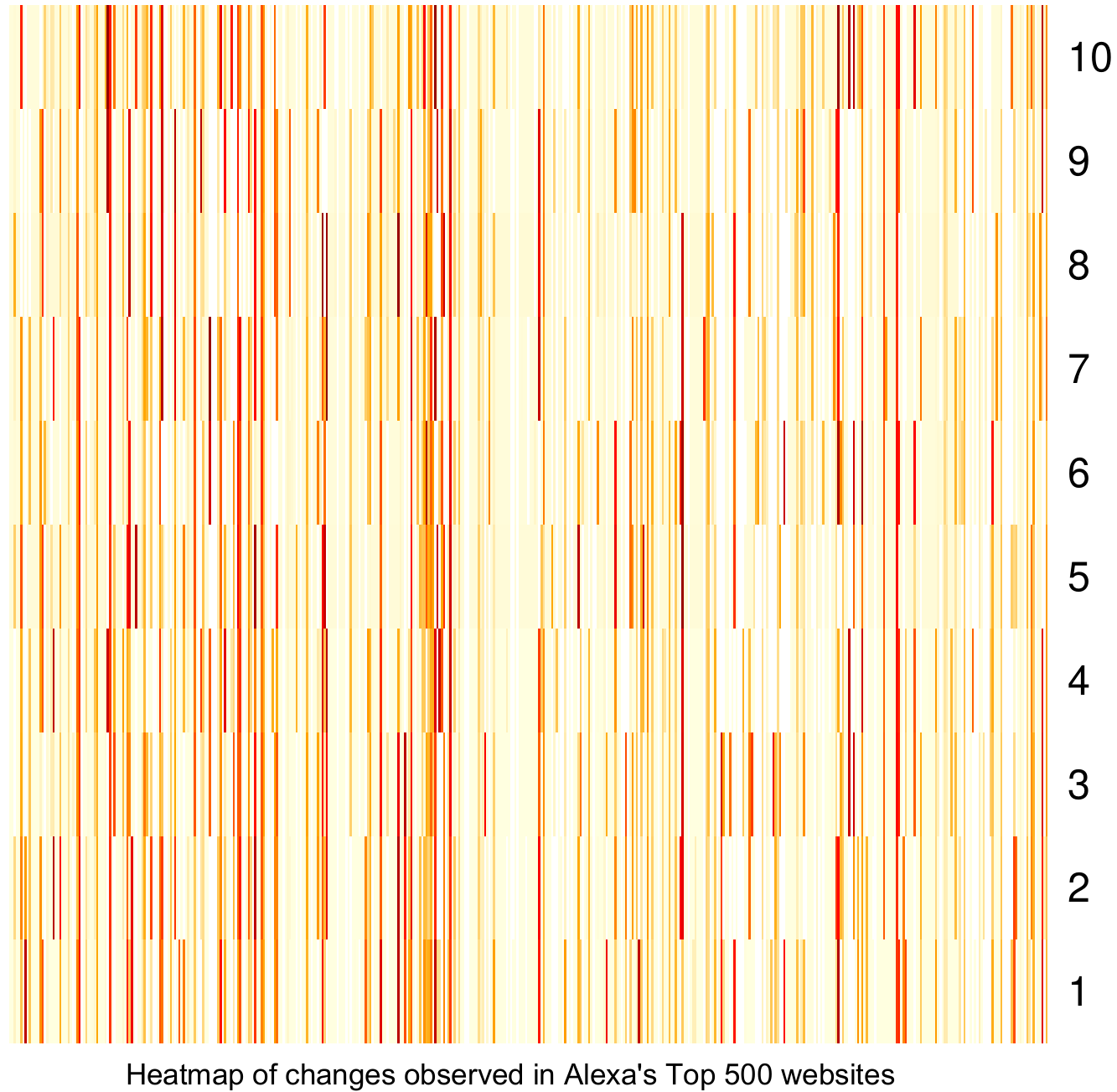}
\vspace{4ex}
\caption{Heatmap for amount of changes observed in Alexa's top 500 website;}
\label{fig:heatmap}
\end{figure}
\begin{figure}[t]
\centering
\includegraphics[trim=30mm 40mm 15mm 5mm, width=0.9\columnwidth]{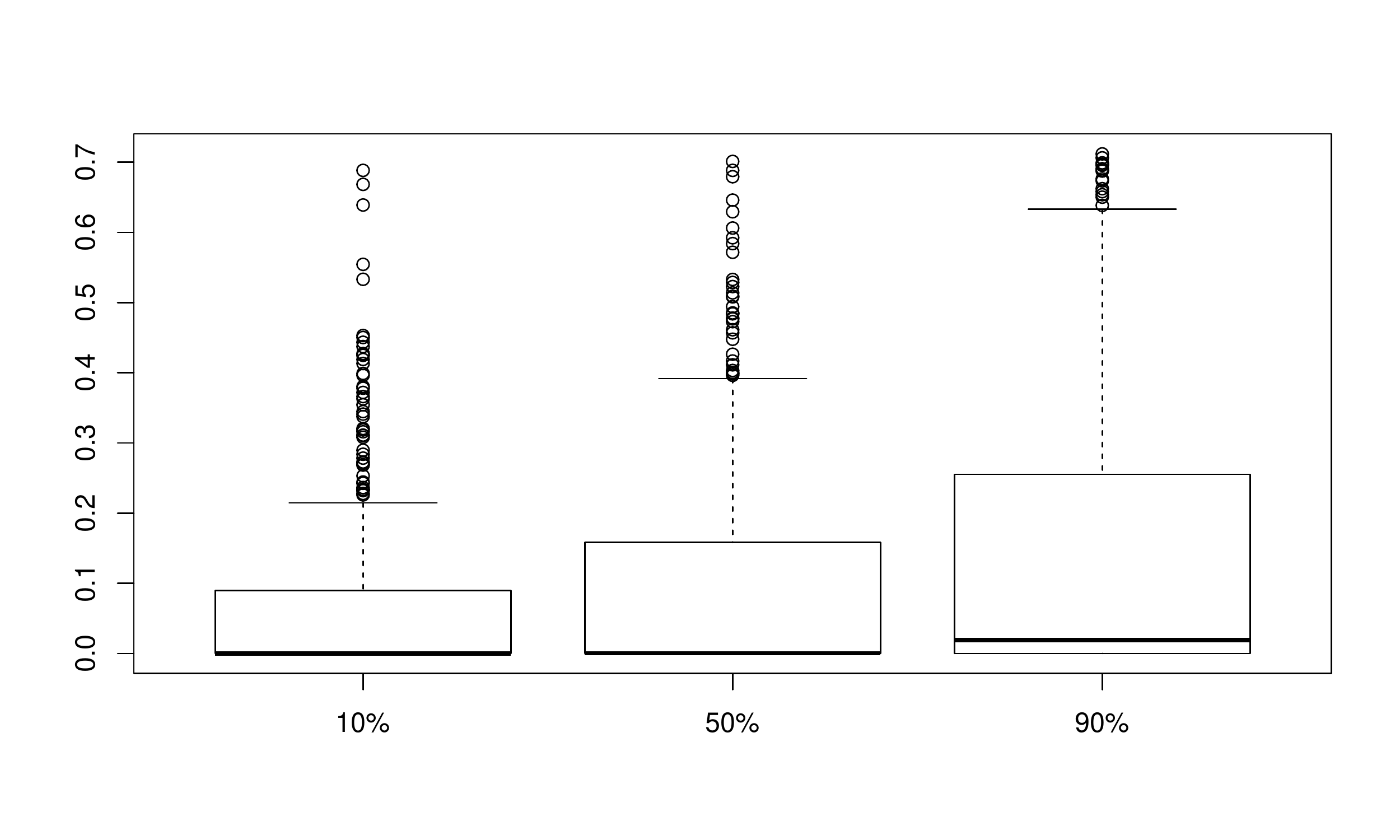}
\vspace{4ex}
\caption{Frequency of visible change in the rendered webpages for Alexa's top 500 websites;}
\label{fig:freq}
\end{figure} 
\noindent {\bf Leveraging the gap in website accesses vs. website changes:}
The access frequency or website hits are normally in measured per second (atleast in multiples of thousands per second for popular sites) where as the frequency with which the websites change is much slower. Even if we consider the same website accessed on different devices separately, broadly there are three types of devices that are used to access websites: Phone, tablet and desktops reducing the access frequency by a factor of 3. These observations suggest the feasibility of harvesting crowd sourced web page snapshots from end user devices which then can be used to provide a very fast interstitial snapshots to other end clients. To quantify the feasibility of crowd sourcing web page snapshots, we repeatedly downloaded Alexa's top 500 websites to analyse frequency of changes in popular websites measuring change in terms of percentage of changed pixels. We took screen shots of all the websites every 2 hours for 24 hours. A look at heat map in Figure~\ref{fig:heatmap} provides the first impression about the change in web pages. Dark red means that almost 100\% of the pixels have changed, bright yellow means that almost no pixels have changed. For each domain, we got a distribution of the percent of changed pixels. From figure~\ref{fig:freq} and looking only at the medians 50\% of time there was substantial change, 50\% there was negligible change, this gives us a new distribution. For that distribution the median is 0, that means that a normally a site doesn't change under normal circumstances from one snapshot to the next for about 2 hours period. The 95th percentile is at 0.4, that means that only 5\% of the sites change 40\% of their pixels every 20 hours. Out of those changed pixels, the change typically includes (i) personalized greetings, (ii) ads, (iii) Moving banners etc. We posit that \sys can {\em leverage this gap in web site access vs. web site change to harvest web pages crowd sourced from end client devices}.

%% file: design.tex
\section{Design Context}
\label{sec:arch}
\begin{center}
\begin{figure*}[t]
\begin{center}
\begin{minipage}{0.40\textwidth}
\includegraphics[width=\textwidth]{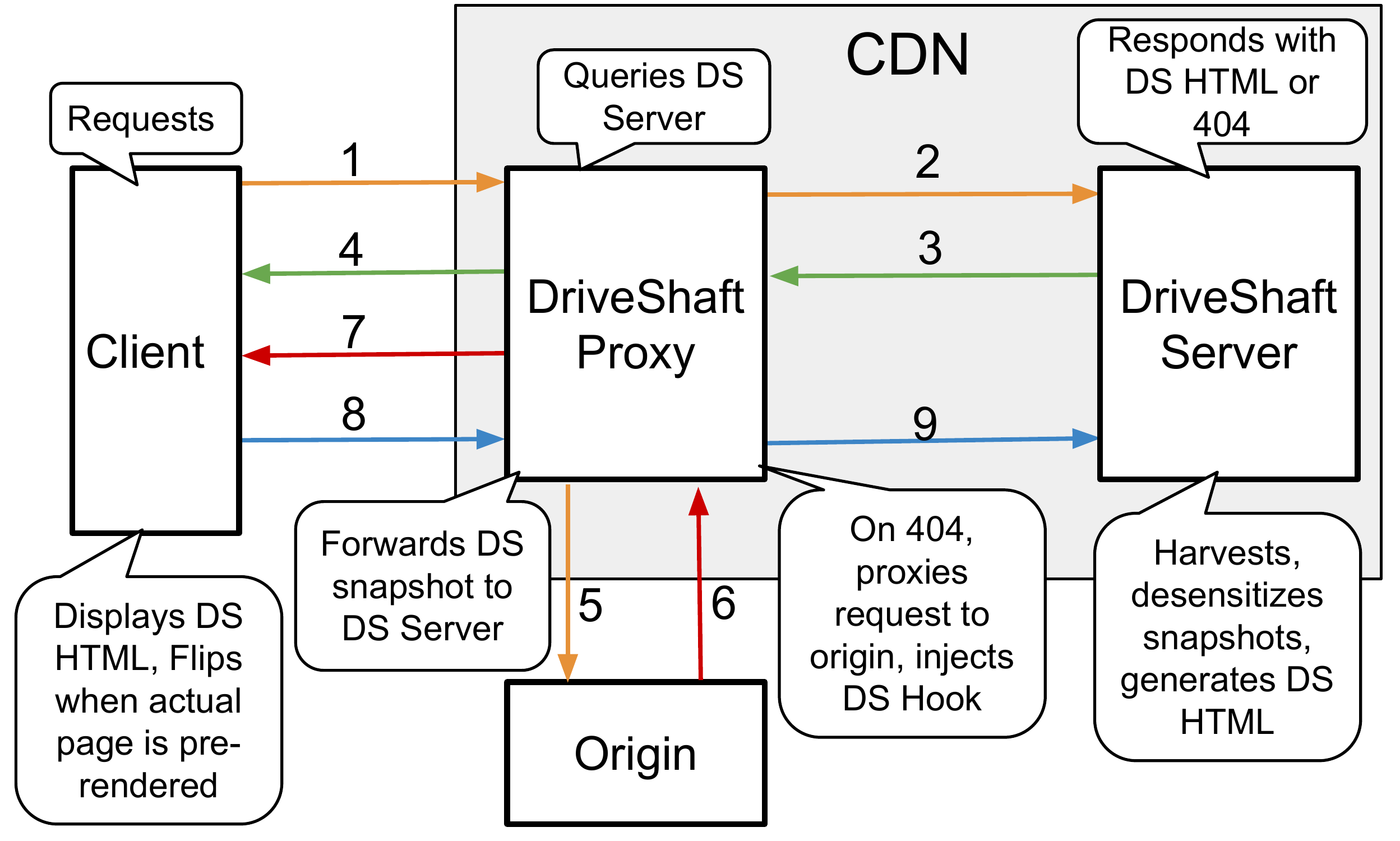}
\end{minipage}
\begin{minipage}{0.58\textwidth}
\includegraphics[width=0.99\textwidth]{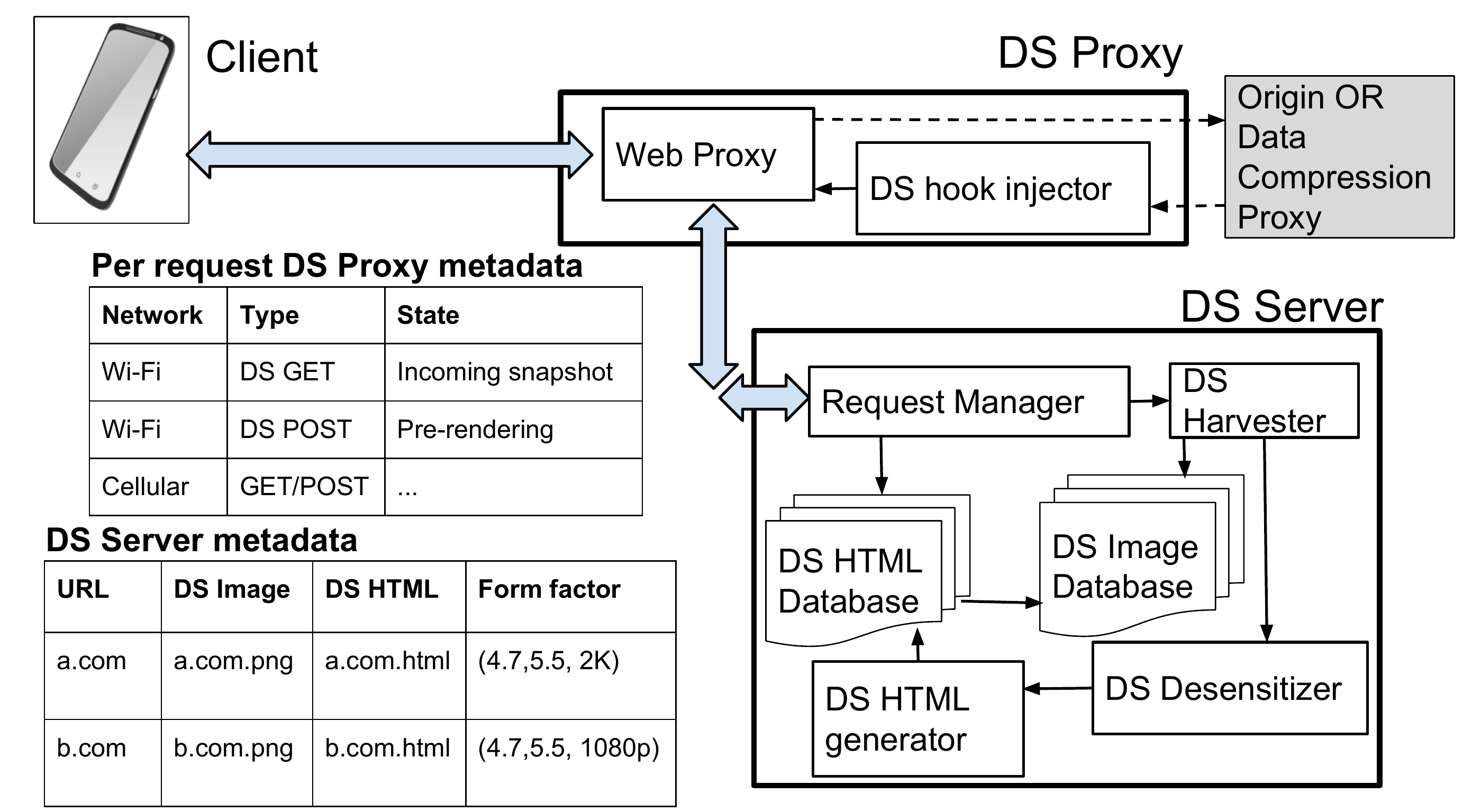}
\end{minipage}
\caption{(i) Showing System Model and overview of \sys operation; (ii) \sys Architecture: Design and interactions of DS modules.}
\label{fig:operation}
\end{center}
\end{figure*}
\end{center}

~\autoref{fig:operation}(i) shows the assumptions for \sys operation derived from how the existing Internet ecosystem \sys operates today. Specifically, \sys spans the following main components:
\begin{tightenumerate}
\item {\it Client}: Mobile devices which access a web page.

\item {\it \sys (DS) Proxy}: Client facing forward proxy. Clients reach it by either DNS-based redirection or by setting up a proxy explicitly on client devices.

\item {\it \sys (DS) Server}: A web server which harvests website snapshots contributed by clients, desensitizes them to remove private information, responds to other clients with appropriate snapshot.

\item {\it Origin}: The web hosting server for the real website.
\end{tightenumerate}

\noindent {\bf Design Goals}

\noindent DS Proxy and DS server are deployed on the nodes or edge servers of content delivery networks (CDNs). Their deployment in CDN is motivated by the fact that most popular websites already use them and having DS modules in CDN nodes doesn't introduce any additional redirection which would adversely affect the time to show visually complete web page otherwise. For \sys
to be practically viable, we zeroed on the following additional design goals:

\noindent {\bf Low Deployment Barrier:} \sys's use in the Internet ecosystem must have very low deployment barriers, i.e., (i) it should not require any additional effort from website developers, (ii) it should be independent and complementary to existing web performance enhancing techniques leveraging them to maximum extent possible and (iii) it should not lead to degradation of web performance in terms of latency.

\noindent {\bf Minimal Footprint:}  It must minimize its footprint i.e., it must touch the least number of elements in existing Internet ecosystem (browsers, protocols, etc.). A concrete example is our choice to avoid adding a new HTML tag to support DS functionality which would require adding associated handling in web browsers. 
A counter example of this is introduction of pre-render tag in Chrome browser to provide functionality of rendering websites in invisible tab -- DS leverages that as described below.

\section{Design}

~\autoref{fig:operation}(i) shows the basic flow of a client request during a website access via \sys. The detailed design, spanning its two main modules -- DS Proxy and   DS Server, and associated metadata -- is shown in~\autoref{fig:operation}(ii), and described in the remainder of this section. 

\subsection{DS Proxy}

DS proxy is the client facing module. It is an enhanced forwarding web proxy which:

\noindent 1. Handles end client requests and if appropriate, gets and serves desensitzed snapshots to end clients.

\noindent 2. Injects a DS hook in the web pages from the origin to capture web page snapshots on end clients.

\noindent 3. Forwards the web page snapshots uploaded by users to DS server for harvesting, desensitization.

To correctly carry out the above mentioned tasks, the DS proxy maintains two types of state: (i) request state: \{clean, pre-rendering, snapshot\} to handle request specific action, e.g., whether to forward a request and/or response to origin or DS server, and (ii) session state: to handle session specific actions, 
e.g., if the connection on device is cellular as opposed to Wi-Fi then don't inject hook to harvest snapshots even if the snap shot isn't available with DS Server. 

The basic operation of the DS proxy can be described as follows: When a client requests a web page using its URL from the origin, it intercepts the call and queries the DS server using the requested URL before forwarding the request to origin. If a snapshot for the requested web page is available, then the DS server returns a DS HTML which contains a snapshot of the requested web page. Otherwise the DS server responds with a 404 HTTP response code, 
and the DS proxy forwards the request to the origin or to a data compression proxy.  
In this manner, the DS proxy works as a simple forwarding proxy, with the exection that it injects a DS hook in the web page. This interaction is shown using orange arrows marked with 1, 2 and 5 in Figure~\ref{fig:operation}(i).

\noindent {{\bf DS hook Injector:}}
\label{sec:clihook}
In order to make the DS operation invisible and reduce the barrier to its roll out, i.e., not requiring developers to make changes to existing web sites, DS proxy injects a DS hook in the HTML served by origin servers on-the-fly. The DS hook is injected only if DS HTML of the requested web page in unavailable or the TTL on the DS HTML has expired. The injection logic needs to make sure that it doesn't break existing functionality such as RUM statistics collection, which may be using the same event, e.g., hooking on load complete event and using navigation APIs to collect performance measurements. 
DS hook adds the following functionality to the HTML page: (i) take a snapshot of a fully rendered web page and (ii) post it to the origin; hence, honoring the same origin policy (SOP). It is important that the post action from client's is only performed after the requested web page finishes loading to minimize its perceivable impact on the web page load time and/or time to show visually complete page. Further, the post action is taken only when the mobile device is connected to Wi-Fi to avoid burdening client's by consuming costly bytes on their data plan. A simplified implementation of DS Hook is provided in Appendix A.

\subsection{DS Sever}

DS server is responsible for: (i) interacting with DS proxy to (a) make available interstitial DS HTML and (b) harvesting the web page snapshots uploaded by user devices, (ii) desensitizing the harvested snapshots to clear them of private information and (iii) generating DS HTML for the websites when number uploaded snapshots cross a threshold and desensitization completes.

To facilitate simple interaction, DS Server exposes a HTTP like interface that DS proxy uses to interact with DS Server. This API is used to lookup, purge and/or add harvested images in DS Image database. Internally, DS server processes desensitization and automatically generates DS HTML which comprises of a click map containing the links as on the requested web page overlaid on its own snapshot with a pre-render hint in link to the requested web page; It also includes functionality to swap DS HTML with the real web page when pre-rendering is completed. The API exposed by DS server and internal modules which handles harvesting, desensitization and automatic generation of DS HTML are discussed below.

\subsubsection{DS Server API}

\noindent {\bf DS GET:} DS GET API is used to retrieve a DS HTML page from the DS server and is built on top of HTTP GET. The client (DS proxy in most cases) uses a HTTP GET request with the original URL in the path of request. For example, in order to fetch DS HTML for www.a.com, client needs to create a HTTP get request to the URL (DS Server)/www.a.com. If an interstitial snapshot for requested web page is available then, DS server returns with a DS HTML or responds with a 404 HTTP response code. This is also shown with orange and green arrows numbered 2 and 3 in Figure~\ref{fig:operation}(i) respectively.

\noindent {\bf DS POST:} DS POST API is designed to facilitate client devices to post a web page snapshot fully rendered page on their browser to the DS server. This is built on top of simple HTTP POST which is automatically performed by the injected DS hook on behalf of end clients. The semantics of DS POST include the necessary information required by DS server to harvest, desensitize, automatically generate a DS HTML in addition to information required maintain correctness while serving DS HTML. The interaction is shown as dark blue arrows marked with 8 and 9 in Figure~\ref{fig:operation}(i). 

DS Post includes the following components:

\noindent {\it DS image:} A snapshot of a web page in a lossless png encoding twice the size of view port on the client device. The usage of png is mandated by the pixel wise comparison employed in desensitization which is not possible with a lossy compression like JPEG. The size of DS Post kept under check by only capturing and uploading snapshots of height twice the size of view port on client device to limit the size of DS Post.

\noindent {\it URL:} A URL of the whose snapshot is posted by clients after web page rendering completion.

\noindent {\it Links:} It contains a set of 5-tuples containing hyperlinks on the web page with co-ordinates in form of \{url,left,top,right,bottom\} used for generating a click map which is overlaid on the snapshot in DS HTML.

\noindent {\it Viewport height:} Size of the view port on client device when the snapshot was taken. It is used by the DS Server to limit the number on links overlaid on the snapshot and determine size of received snapshot.

\noindent {\it Form factor:} It contains a set of 3-tuple or 4-tuple containing the resolution of client device and its physical form factor (if available) of its screen measured in inches. For standard resolutions a single string is sufficient but for non-standard resolution of the device, we require both width and height in terms of pixels e.g., \{4.7, 5.5, 1080p\} or \{4.7, 5.5, 1320,1024\}. This is then used to create a appropriate device specific DS HTMLs.

\noindent {\it User Agent:} It is not a part of POST per say but is already available in DS Post. This is extracted and is later used to serve the correct snapshots to devices supporting different operating system e.g. iOS vs. Android.

\subsubsection{DS Harvester}

The DS Harvester is responsible for storing the incoming snapshots from clients as part of DS POST to DS image database. For each received snapshot, it maintains time stamp, the URL, the information in DS post. Specifically, it maintains a table for snapshots uniquely identified by a 3-tuple \{time stamp ,requested URL, form factor tuple\}. However, at this time DS HTML is not available for servicing clients till a pre-defined number of web page snapshots have been received by DS Harvester. DS harvester uses this tuple for the following purposes: (i) To trigger the desensitization process when the number of received snapshots reach the desensitization threshold, (ii) To maintain freshness of desensitized image by periodically comparing the time stamp of latest snapshot received with the time stamp at which the most recent desensitized image was generated and triggering desensitization process when required and (iii) To ensure generation of unique form factor specific DS HTML for devices. 

\subsubsection{DS Desensitizer}

\begin{figure}[t]
\centering
\includegraphics[trim=30mm 25mm 15mm 15mm, width=0.9\columnwidth]{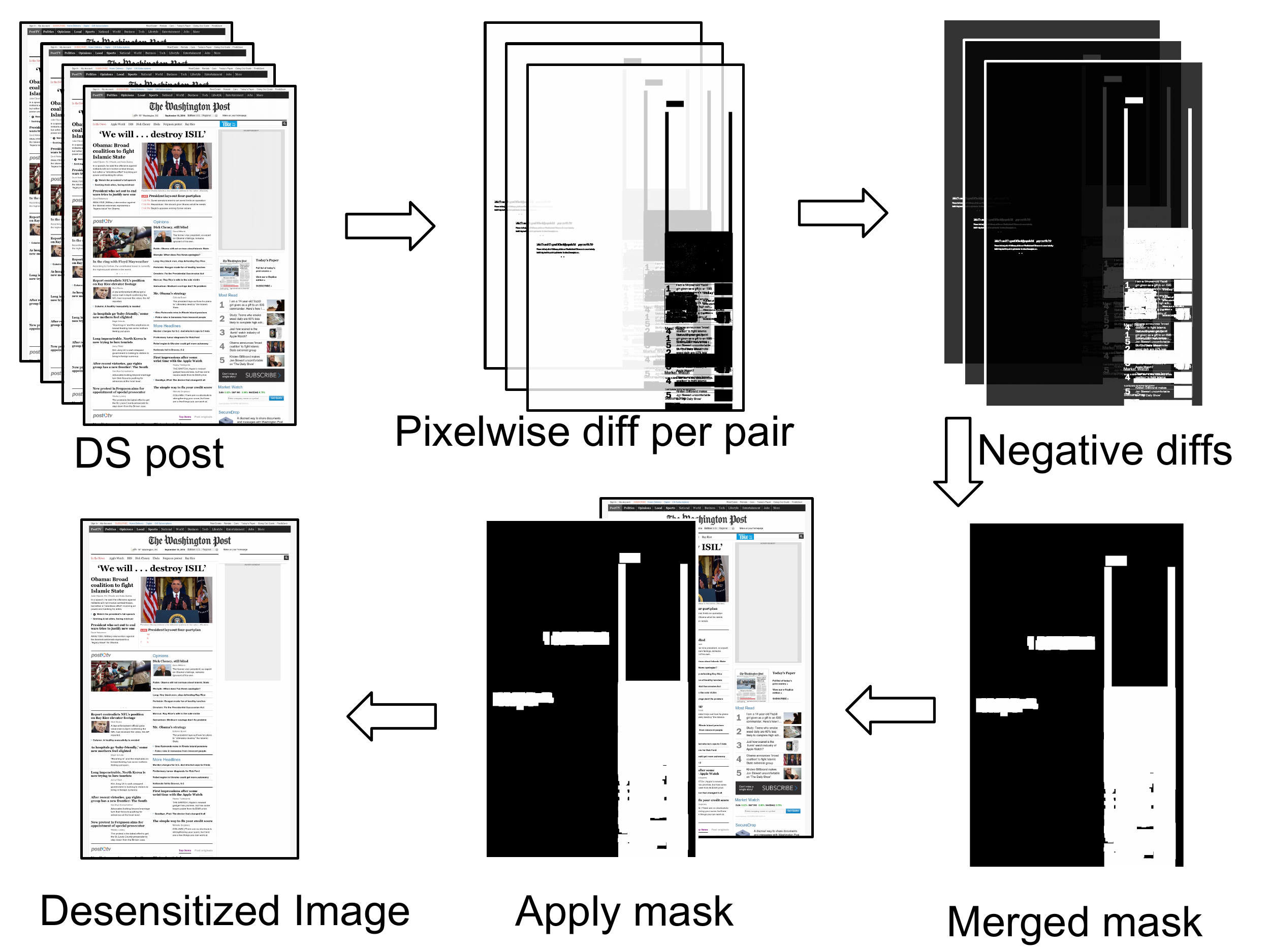}
\vspace{4ex}
\caption{ Showing steps in \sys's desensitization of web page snapshots.}
\label{fig:desen}
\end{figure}

DS Desensitizer module is responsible for ensuring that the images used in DS HTML to service other clients are clear of any sensitive and/or private information e.g., the snapshots posted by clients may include greetings containing names of end user, ads that are seen by users when they visited the page etc. The desensitization process is illustrated in Figure~\ref{fig:desen} using a real run shows the different images produced during desensitization process. Currently, it involves pixel wise comparison of multiple snapshots for the same \{URL, form factor\} to create a mask from each image, then taking a negative of those masks to create a single mask covering maximum difference in the snapshots. The mask is then overlaid on the original snapshots to create a desensitized web page snapshot. This process ensures that the areas containing information that is different in the snapshots (hence, potentially sensitive) posted is blanked out in the snapshots which are served to the clients. The desensitization removes user specific information, ads and any other dynamic content present on the web page snapshot. This process is triggered asynchronously by DS harvester because computational intensive nature if image processing tasks. Based on our observations, it takes considerable amount of time varying from a few seconds to a few minutes to complete. We used three snapshots for faster experimentation but this can be increased based on individual websites and the degree of sensitive information on the web page. Better image comparison algorithms such as keypoint matching, histogram method etc., and sensitive text removal based on optical character recognition (OCR). 
 
\subsubsection{DS HTML generator}

The desensitized image, the links and their co-ordinates on the web page are used to generate appropriate DS HTML. DS HTML generator module is responsible for automatically generate HTML pages (referred as DS HTML) containing the snapshot created as a result for desensitization process uniquely identified by \{URL, form factor\}. DS HTMLs are then used to service the client's requests via DS proxy. In addition to the link to web site snapshot, DS HTML contains a pre-render link to the original requested URL, an overlay of click-map containing the links same as the original web page and functionality to a redirect to the real web page when pre-rendering process is completed. DS HTML adds a token that allows DS proxy proxy doesn't forward pre-render request to the DS server but directly forwards those requests to origin in order to fetch and serve the original web page without any changes and/or injecting DS hook in case it is  required. The detection of pre-render request is facilitated by addition of an additional parameter in path of the requested web page in the pre-render link which is added to the DS HTML during its generation process. On detection of this additional parameter, DS proxy acts as a simple forwarding proxy. While generating DS HTML, the click-able links are overlaid over web page snapshot using "imagemap" HTML tag and the stored co-ordinates received in DS Post. 

\section{Interesting Design Considerations}
In designing \sys there are a number of important considerations that we found interesting. We discuss those in this section.
\subsection{Handling Devices' Form Factors}

Difference in device resolution, physical dimensions and/or orientation of device while browsing leads to different snapshots being posted from end client devices. DS server uses the information in DS post to store multiple copies of DS images and hence, DS HTML for devices of different form factors. However, it is unreasonable number of combinations given that there are so many different form factors for android devices. This is more severe for native apps where UI widgets may differ in size compared to browsing mobile websites because (i) using higher resolution pictures to devices supporting lower resolution but not vice-versa and (ii) most popular websites including those employing responsive design support only three modes (i) phone, (ii) tablet and (iii) Desktop which is also evident from the three user-agent types supported by popular browsers e.g., chrome, Firefox etc.~\cite{useragent}. There are interesting trade-offs here e.g., we can reduce the number of snapshots to be maintained at edge servers by only maintaining snapshots for high resolution images and downscaling them to three form factors which reduces pressure on clients i.e., the number of clients posts by limiting posts to high resolution phones but that (i) puts pressure on lower resolution phones to fetch a larger image that they can really use and (ii) affect the freshness of DS snapshots. We posit that this is tunable and specific to websites and would basis of negotiation between CDNs and content providers on QoS expected from DS service.

\subsection{Handling Transient Unavailability of Dynamic and Interactive content}
 
Using DS leads to a transient period when there is no availability of dynamic content (e.g., moving banners, ads, etc.) and interactive content (e.g., as text fields etc.) from web pages i.e., in the time period between when DS HTML is rendered interstitially and the actual requested web page is being pre-rendered in background. However, the duration of this phase depends on difference between the real web page load time and the DS HTML load time. During this time user cannot interact with the web page anyways so in practice DS has very small impact on interactivity. In fact, it complements the existing web performance techniques by masking the delay in web page loading but still being unable to provide visual response during the interstitial DS HTML and the real webpage flip. This is exactly the reason, we think that DS can also hide the latency in time to show visually complete web page. From a deployment perspective, with co-operation between a CDN providing DS service and content provider, the dynamic content can also be populated on DS images on-the-fly. DS prototype supports this by logging the rectangle co-ordinates of the desensitized parts on the web page and then, providing hooks to add can add the content (e.g., ads, banners etc.) at the location of the desensitized part of web page snapshot. The content however, needs to be made available to DS servers via an API or as content by origin.

\subsection{Handling Personalization}

DS desensitization process leads to masking of different pixels resulting from comparison of the snapshots posted by different end client devices. However, most modern websites provide a personalized response even without logging in using HTTP cookies stored on client devices when the user revisits. Similar to those, a mask for personalized content from client's cache is overlaid on the served DS snapshot before displaying it on end user device. However, this can introduce additional delay in displaying the DS snapshot highlighting a trade-off in personalization vs. latency. The extent to which this personalization happens is tunable and is another trade-off that has negotiated by CDNs and web site developers. 

\subsection{Handling Privacy Concerns}

Since, DS requires user to upload snapshots of the websites they view on their mobile phones, one deployment concern with DS is managing user privacy. The concerns arise in two steps: (i) Concerning harvesting of mobile web snapshots, we argue that uploading snapshots of web pages that, in fact, are served by CDNs themselves does not degrade end user privacy in any real sense. Also, since CDNs handle end user content securely on its servers on behalf of service providers under strict agreements, user privacy is not compromised in any meaningful ways by harvesting them back at CDN nodes. The level of privacy offered by DS is no inferior than Google's data compression proxy which has all the user requests and responses to compress the content on their servers and transform them to reduce data costs at end client devices. (ii) Another privacy concern that may arise is from false sharing of web page snapshots among different end users and/or geographically separate nodes. A simple approach of masking different pixels in snapshots of the same web page is very effective in combating false sharing i.e., in most cases, the sensitive information will be different for different users. If an information is not different for many users, then it is highly unlikely that it is private. We have another tunable parameter here i.e., by increasing the desensitization threshold i.e., the number of snapshots posted by client devices to make available a desensitized snapshot to end users, we can decrease the likelihood of sensitive information being leaked because it is more probable that the portions of web pages which are different among different users will be blurred during desensitization process. Additionally, more advanced desensitization techniques like using optical character recognition to read text in the posted snapshot and using regular expression to identify sensitive info e.g., credit card number, social security number etc. and adding that portion to mask that is used for blurring, it can removed during desensitization process.

%% file: impl.tex
\section{Implementation}
\label{sec:impl}

The DS prototype is implemented using standard python libraries, peewee - ORM and MySQLdb are used to implement DS harvester. DS Desensitizer is implemented using imagemagicK - image processing tool kit. Since, DS's design goal was to ensure that it doesn't require any change for web sites and in browsers, the DS Hook is implemented as a JavaScript which is supported by all browsers. DS HTML generation is done using simple text manipulation and using HTML tag image map to overlay links on the snapshots. We faced a few counter-intuitive implementation challenges in prototyping this seemingly simple system enumerated below:

\noindent 1. To handle pre-render request from a DS HTML page correctly at DS proxy, DS proxy needs to maintain an extra state. This is implemented as adding an extra pre-render to the actual URL request which is them removed by DS proxy. But, this can problematic in cases where actual pre-render request never arrives from client because may be client closed the browser or any other error. Current prototype handles this by clearing this proxy state after a time out.

\noindent 2. DS Hook injector implementation is not simple as the format for the HTMLs varies greatly e.g., a simple approach to replay <head> tag wouldn't work if there are multiple head tags or simply adding adding code to carry out DS POST on loadComplete event might be incompatible with any other existing functionality e.g., monitoring web performance using RUM stats etc. So, DS Hook injection to the proxied HTML can be tricky to implementation correctly for all websites. For CDNs, this means that DS Hook would have be customized and/or tested for every website, a practice that is already in place for front end optimization(FEO) techniques.

\noindent 3. Client's web cache is polluted while browsing internet via DS because the responses for real request are DS HTMLs and the responses to pre-render added URLs are the actual web responses. We posit that this would be solved as DS gets deployed to more and more devices but for now this remains part of our future work to fix client's web cache.
\noindent 4. Server side redirects and cookies in requests for the same URL create implementation and/or deployment issues. This was observed for a small number of websites and can be handled on per web site basis in DS harvester to ensure the their correctness.

%% file: eval.tex
\section{Evaluation}
\label{sec:eval}

\begin{center}
\begin{figure*}[t]
\begin{center}
\begin{minipage}{0.33\textwidth}
\includegraphics[width=\textwidth]{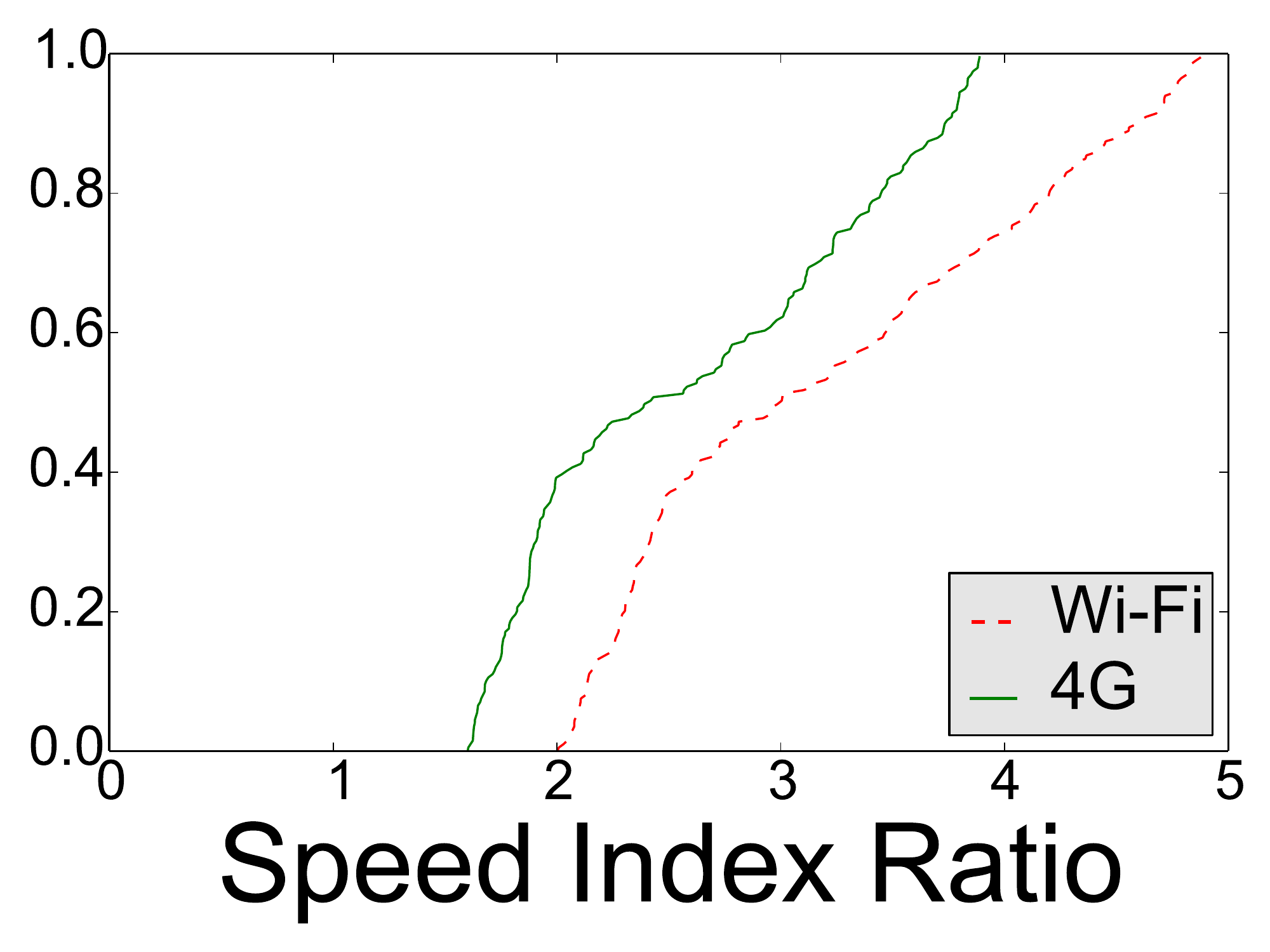}
\end{minipage}
\begin{minipage}{0.33\textwidth}
\includegraphics[width=\textwidth]{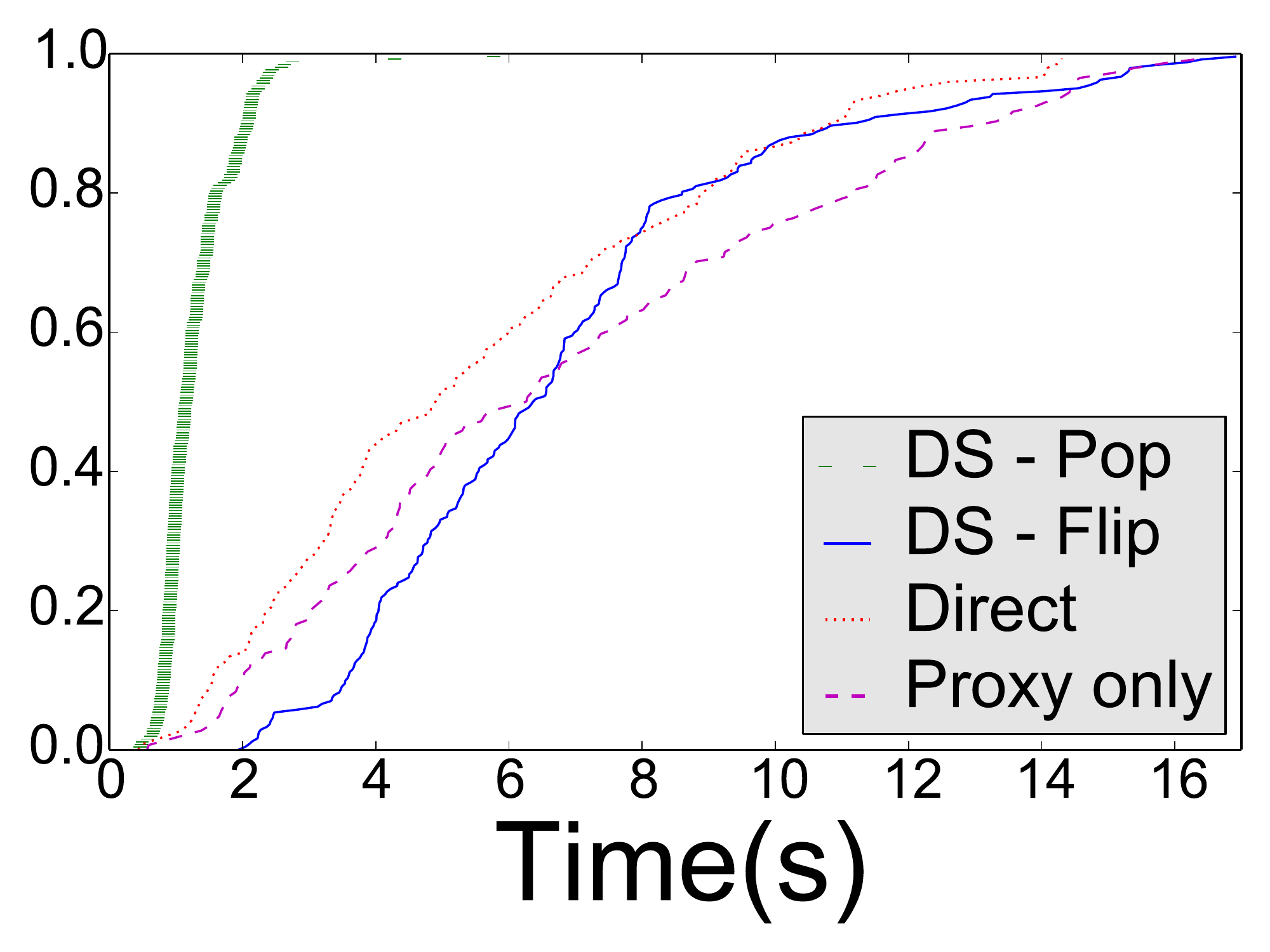}
\end{minipage}
\begin{minipage}{0.33\textwidth}
\includegraphics[width=\textwidth]{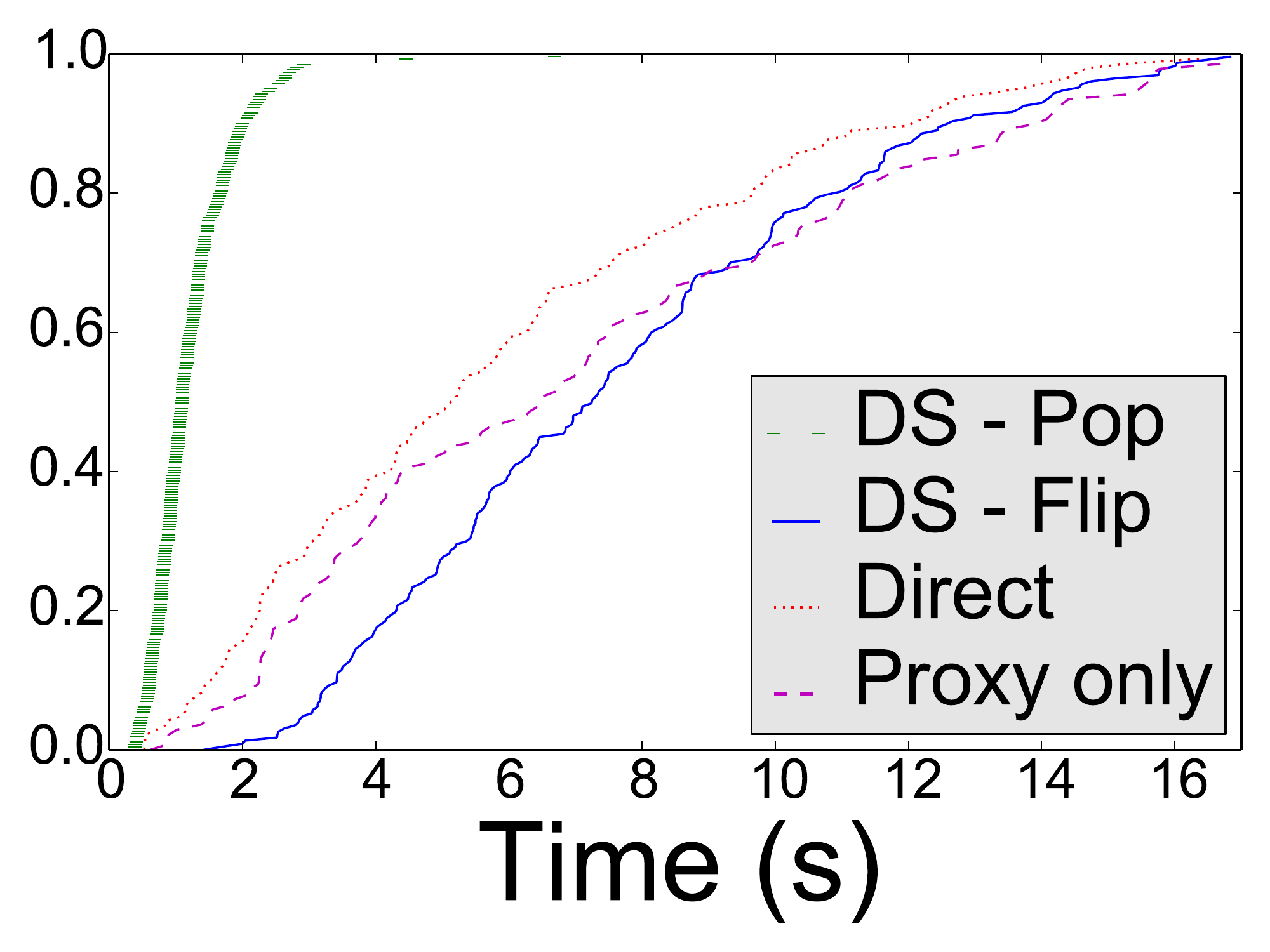}
\end{minipage}
\caption{\small DriveShaft Performance using Moto X: (i) CDF of Speed index ratio over Wi-Fi and 4G-LTE; CDF comparing timing of web page load time when accessing internet directly, using a simple proxy, using data compression proxy and load time for DS HTML as well as the DS Flip over  (ii) Wi-Fi and (iii) 4G LTE.}
\label{fig:perf}
\end{center}
\end{figure*}
\end{center}

\noindent {\bf Experimental Setup:}
The end to end set-up for evaluating DS comprised of two devices i.e., (i) Motorola Moto X and (ii) Motorola Moto E as representative of a high end and a mid-range mobile phones. The Internet connection used was (i) high speed Internet connection over 802.11n wifi connection and (ii) commercially available 4G LTE. The DS proxy and DS server are deployed on a local test bed on our local openstack based cloud providing us with VMs with following configuration: Ubuntu 14.04 VM on local cloud test bed: 4 vCPUs, 16GB Memory mimicking CDN nodes.  

\noindent {\bf Workload:}
We started with 300+ sites but finally ended up with a set of 200 websites from Alexa's list which worked correctly with DriveShaft. The selection was based on one main criteria i.e., the websites must not require HTTPS connection. This decision to choose only HTTP websites for end to end evaluation was to clearly highlight the benefit of DS by eliminating impact of handling split TCP in proxy or TLS. However, we made sure that (i) the websites varied considerably in their complexity, (ii) spanned across numerous categories like news, online stores etc. and (iii) did not contain only a few elements in their layout e.g., search box, login etc.

\noindent {\bf Measurement and metrics:}
The measurements were carried out using automated measurements defined in Chromium Telemetry which allows running a set of performance measurement against a web page set. We repeated our experiments multiple times on live sites to ensure that our results reflect the observations in real world scenarios and iron out any irregularities in the measured performance. For comparison with newer browsers, we had to manually run experiments because Telemetry integration with them was too difficult resulting in a reduced subset of 50 websites.
\noindent We use the following terms in this section to explain the experimental results: 

\noindent {\small {\bf Speed Index(SI):}} The Speed Index~\cite{speedindex} is the average time at which visible parts of the page are displayed.

\noindent {\small {\bf Timing:}} Web page load time measured from performance.timing.navigationStart until the completion time of a layout after the window.load event for the following:

\noindent {\small {\it DS Pop:}} Time to load and render DS HTML.

\noindent {\small {\it DS Flip:}} Time to finish DS tiggered pre-rendering. At this point, user's view is flipped to real web page.

\noindent {\small {\it Direct:}} Page load time when accessing directly.

\noindent {\small {\it Proxy:}} Page load time via proxy (base for DS proxy). 

\begin{table}[t]
\begin{tabular}{|p{1.5cm}|p{0.55cm}|p{0.55cm}|p{0.70cm}|p{0.70cm}|p{0.70cm}|p{0.70cm}|}
\hline
\multirow{2}{*}{Time(ms)} & \multicolumn{2}{c}{DS Pop} & \multicolumn{2}{c}{DS Flip} & \multicolumn{2}{c|}{Direct} \\
 & X & E & X & E & X & E \\
 \hline
 Min. & 379 & 456 & 1473 & 1937 & 436 & 1236 \\ 
 1st Quant. & 719 & 935 & 3674 & 4602 & 2870 & 6422 \\
 Median & 945 & 1148 & 5250 & 6656 & 5093 & 8040 \\
 Mean & 1120 & 1296 & 6839 & 7984 & 7397 & 10350 \\
 3rd Quant. & 1429 & 1489 & 7862 & 8599 & 8685 & 13100 \\
 Max. & 4004 & 5769 & 53830 & 45330 & 67220 & 67230 \\ 
 \hline
\end{tabular}
\caption{\it DriveShaft time line for 200 popular websites on a low end smart phone i.e., Motorola Moto E vs. a high end smart phone i.e., Motorola Moto X}
\label{tab:perf2}
\end{table}
\begin{center}
\begin{figure*}[t]
\begin{center}
\begin{minipage}{0.33\textwidth}
\includegraphics[width=\textwidth]{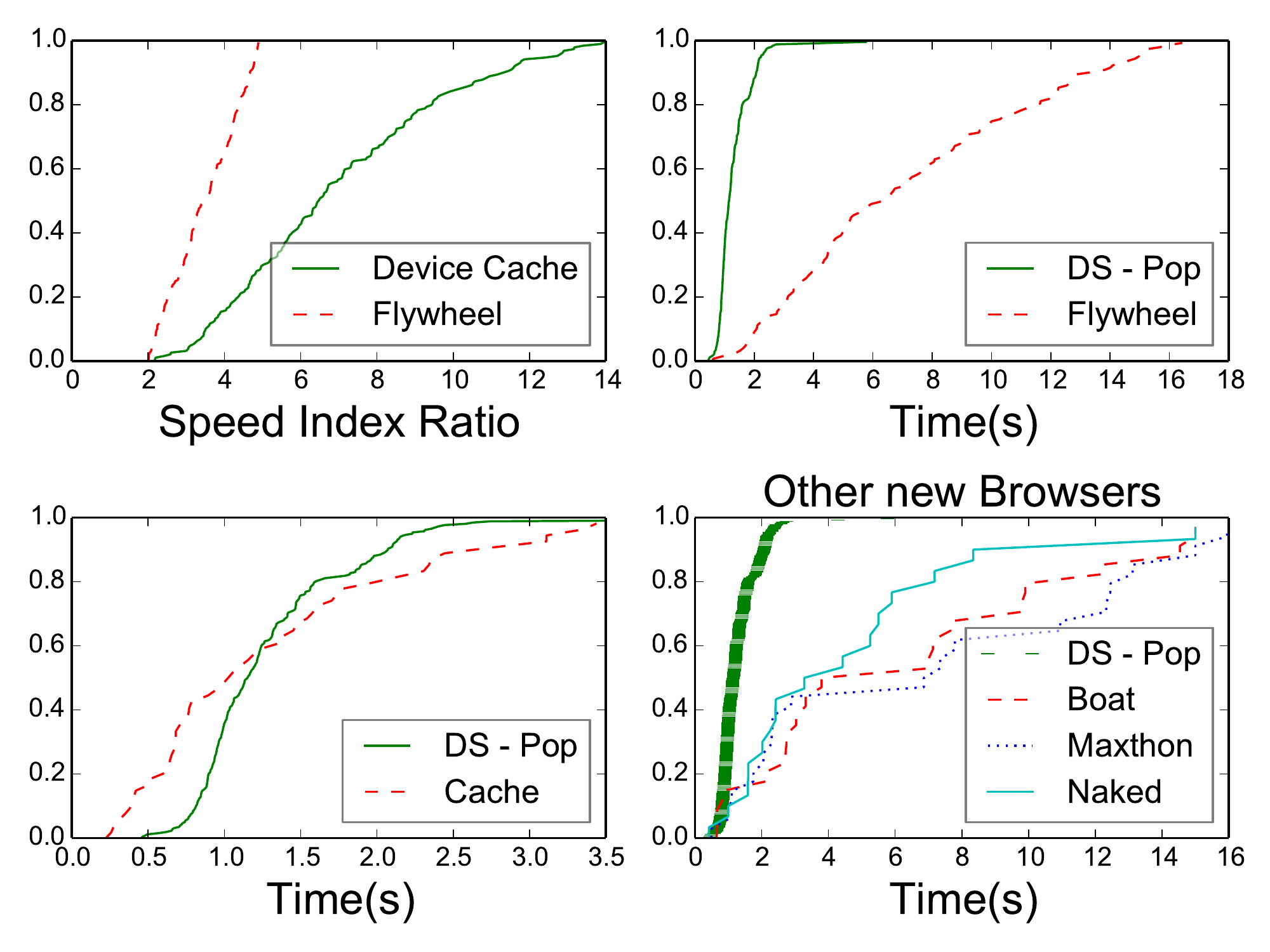}
\end{minipage}
\begin{minipage}{0.33\textwidth}
\begin{tabular}{|p{1.6cm}|p{1cm}|p{1.35cm}| }
\hline
\%age & Median & 90th \%ile \\ 
\hline
Unchanged & 84.1 & 86.4 \\ 
\hline
Masked & 19.2  & 29.5 \\ 
\hline
False -ve & 0.8 & 1.1 \\ 
\hline
False +ve & 12.1 & 15.6 \\ 
\hline
Discards & 9.7 & 11.7 \\ 
\hline
\end{tabular}
\end{minipage}
\begin{minipage}{0.33\textwidth}
\includegraphics[width=\textwidth]{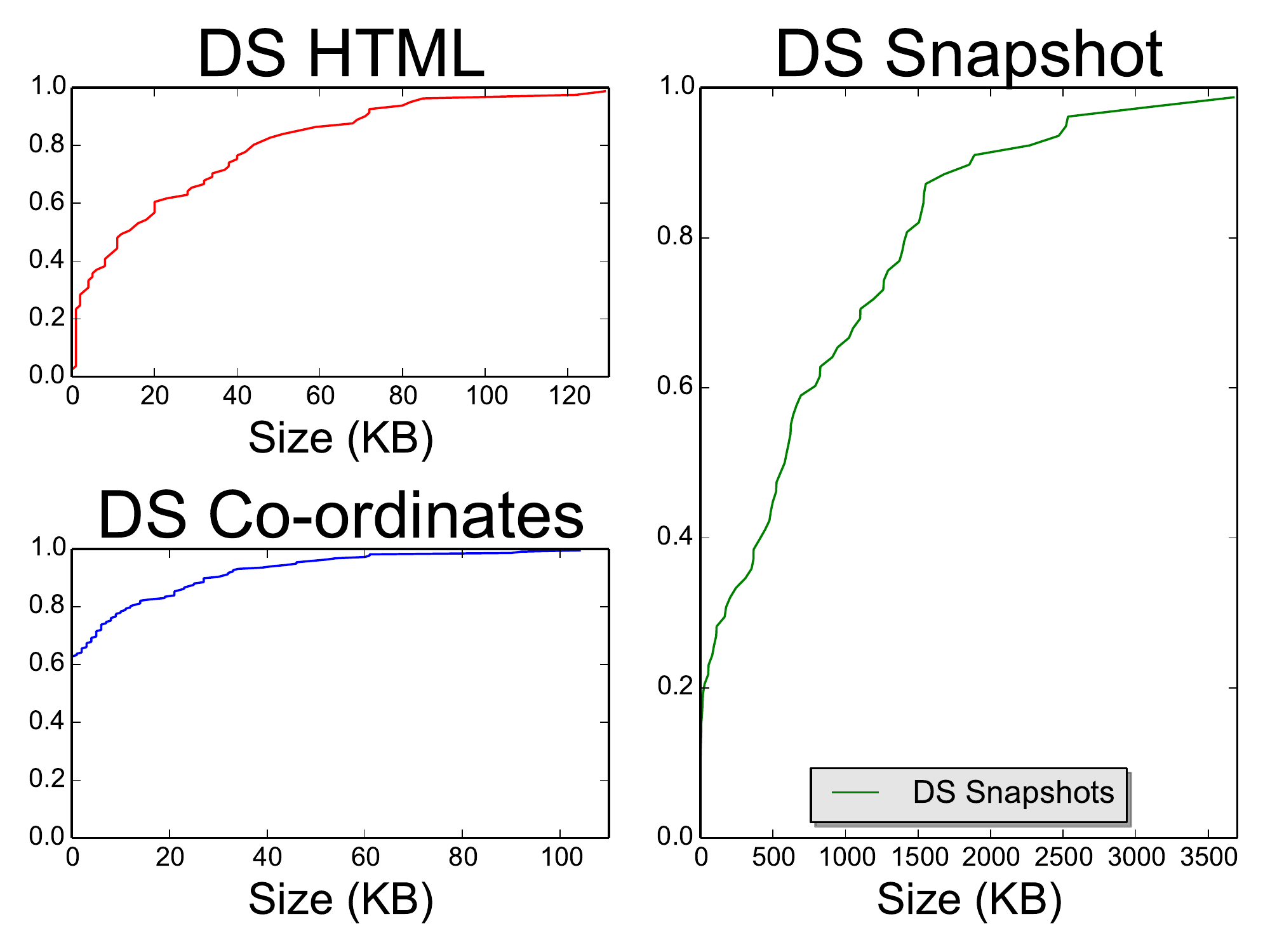}
\end{minipage}
\caption{\small Showing (i) (a) Speed index ratio of DriveShaft to FlyWheel - Google's data compression proxy and device's local web cache, (b) comparing web page load time: Flywheel vs. DriveShaft; (c) Comparing web page load time: DriveShaft vs. Device's local web cache and (d) DriveShaft web page load time with newer browsers: Boat, Maxthon and Naked; (ii) Table showing empirical observations for desensitization process in terms of percent of pixels that remained unchanged, masked, wrongly unchanged i.e., false negatives, wrongly masked i.e., false positives and percent of snapshots that caused discarding of harvested snapshots due to too much change; (iii) DriveShaft extra bytes on wire using CDF of the sizes of DS Snapshots, co-ordinates data and generated DS HTML which will be used to serve other clients.}
\label{fig:comp}
\end{center}
\end{figure*}
\end{center}
\begin{center}
\begin{figure*}[t]
\begin{center}
\begin{minipage}{0.33\textwidth}
\includegraphics[width=\textwidth]{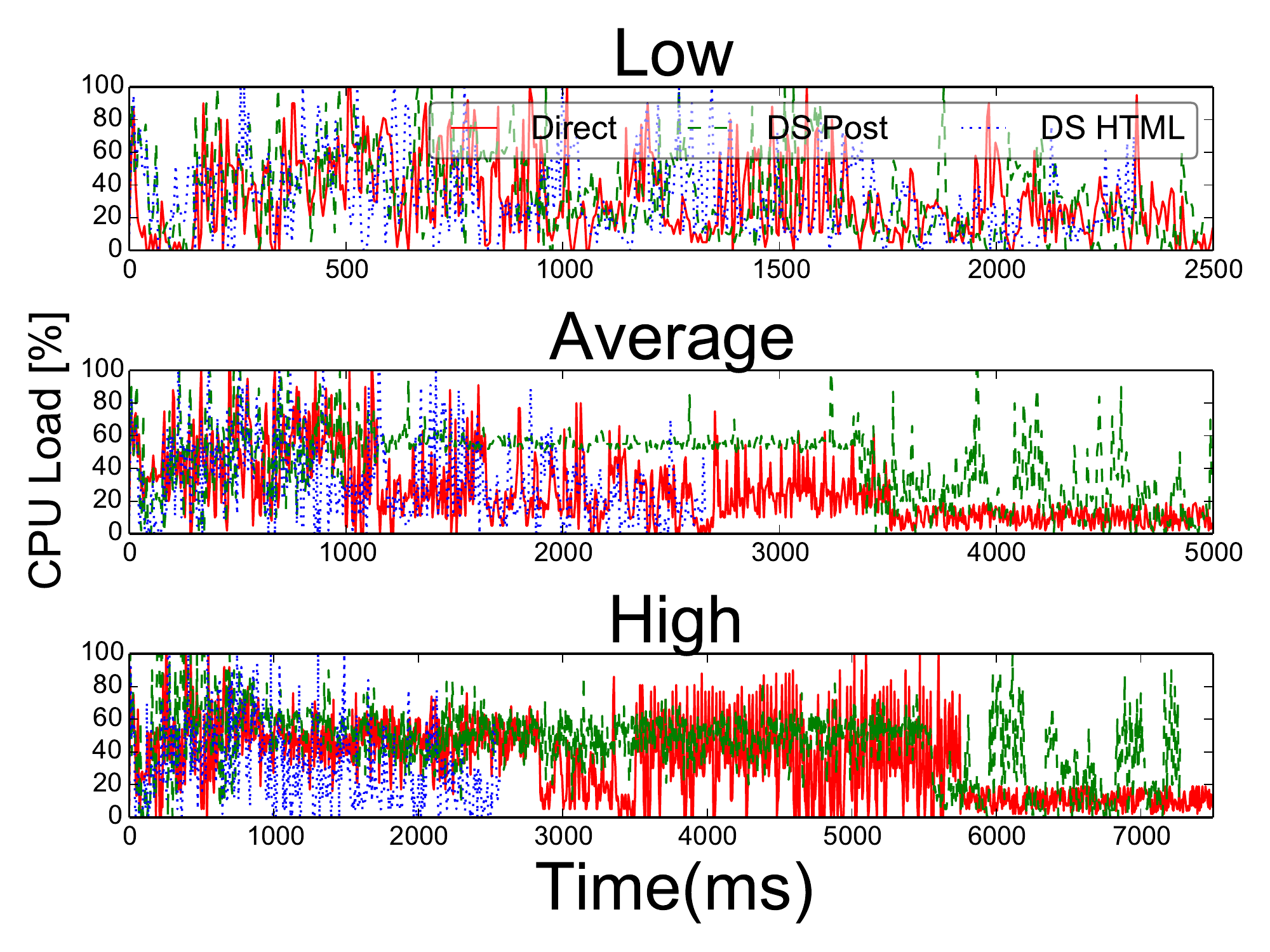}
\end{minipage}
\begin{minipage}{0.33\textwidth}
\includegraphics[width=\textwidth]{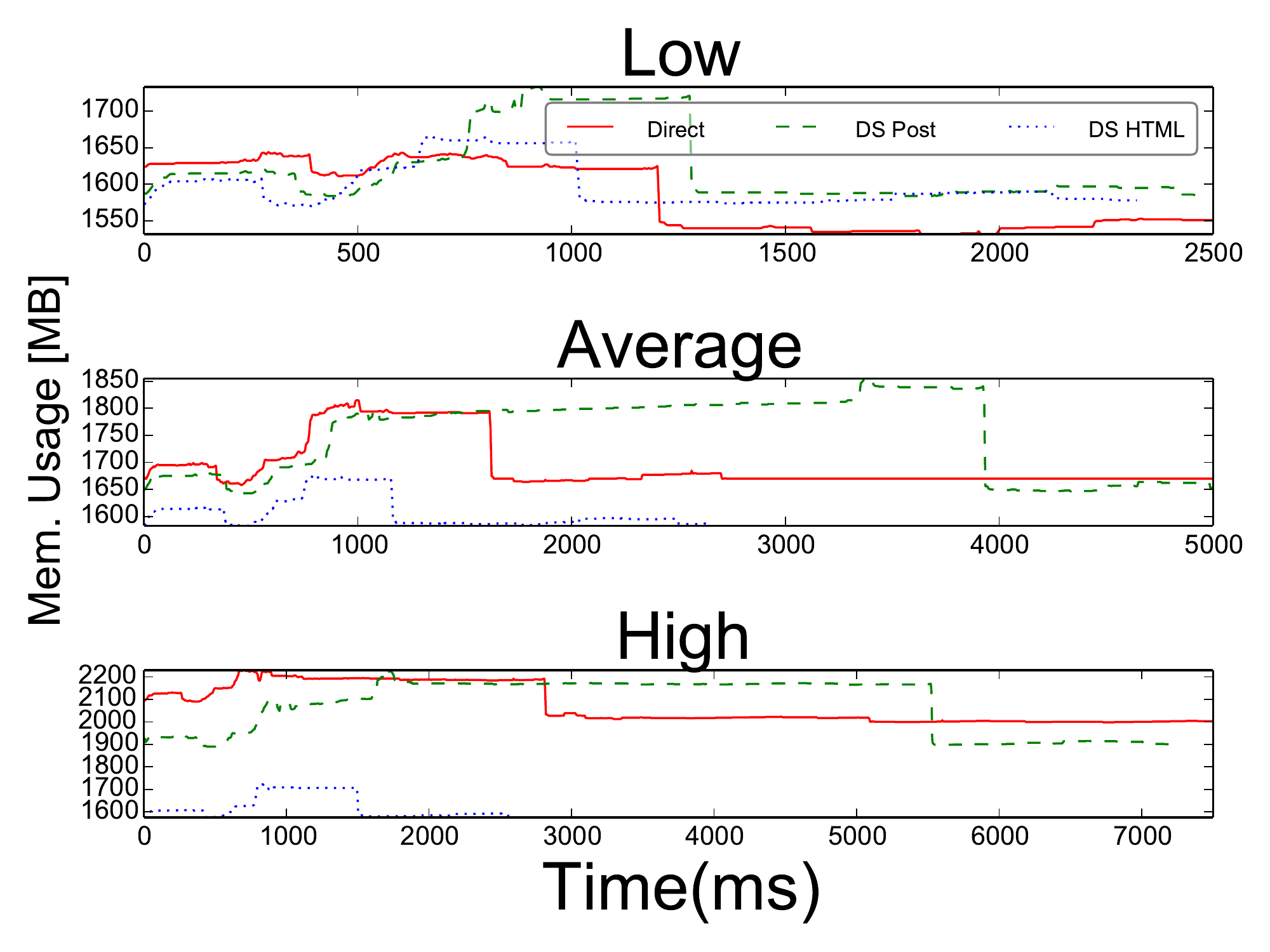}
\end{minipage}
\begin{minipage}{0.33\textwidth}
\includegraphics[width=\textwidth]{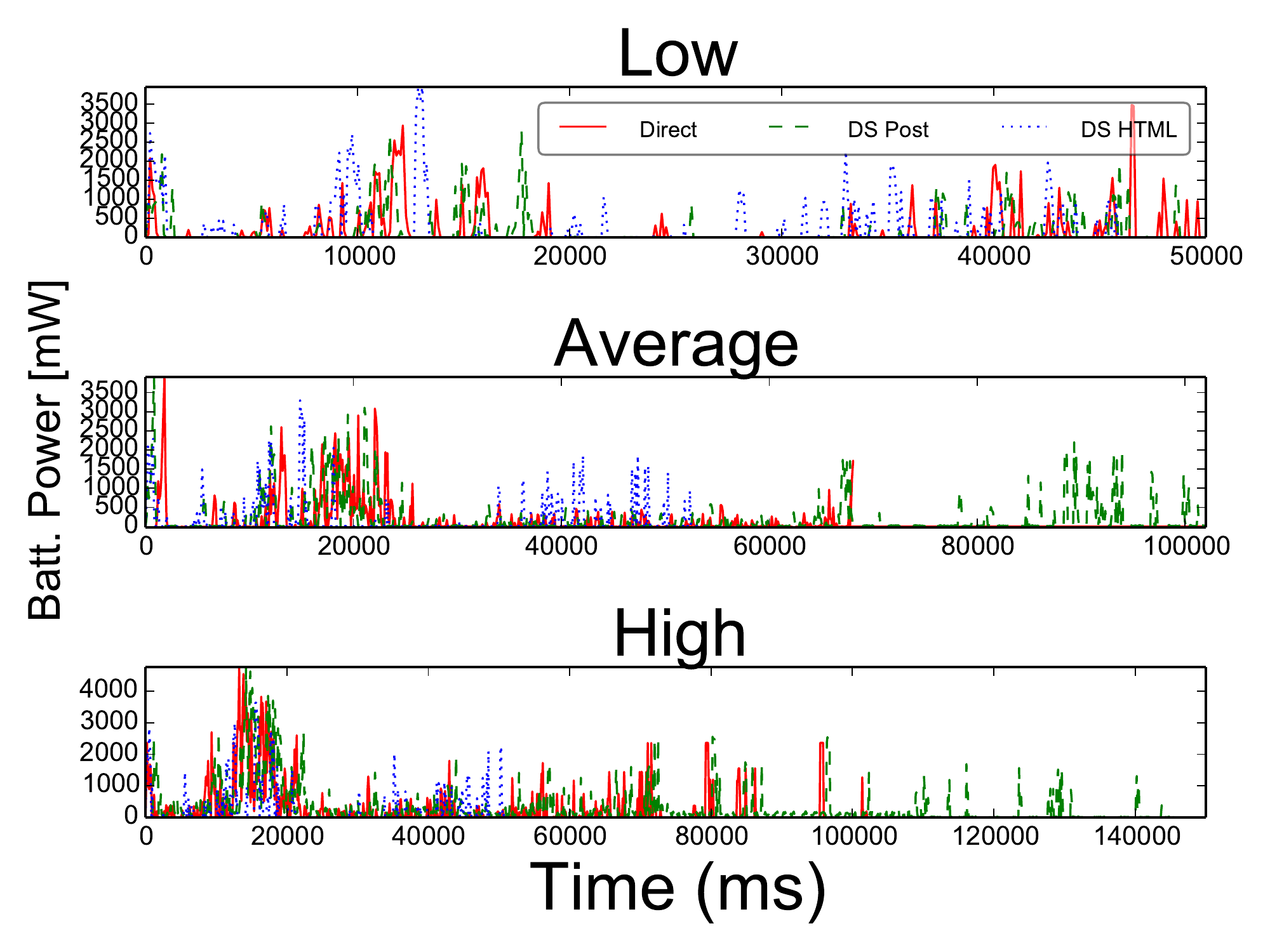}
\end{minipage}
\caption{\small Showing DriveShaft overhead (i) CPU; (ii) Memory; (iii) power on client device using web pages with low, average and high complexity.}
\label{fig:over}
\end{center}
\end{figure*}
\end{center}

In evaluating DriveShaft, we seek the answers to the following questions:

\noindent {\em Q: Does DS lead to faster appearance of web page loading or fills user screens faster?}

We measured speed index (SI) when accessing a website directly form the Internet vs. via DS proxy over Wi-Fi and Cellular 4G network using Moto-X smart phone. Then, we calculated SI ratio as ratio of SI measured when using DS to SI measured when directly accessing website. Figure~\ref{fig:perf}(i) shows the CDF of measured SI ratios. The improvements in speed index range from 2 to 5 over Wi-Fi and from 1.6 to 3.9 over 4G. In favor of brevity, we only include plots of SI ratio to highlight the relative improvements but it is important that we report that the measured absolute speed index when using DS proxy was consistently less than 1000. This is simply due to the fact that DS HTML containing the snapshot can be fetched and rendered very fast i.e., only one request-response cycle to quickly fill the screen resulting in lower speed index regardless of the website complexity or the time a website actually takes to pre-render in the background. This also shows that DS effectively decouples perceived web page performance from its structural complexity and without requiring any change by website developers or end users achieving its goal of low deployment barrier. {\em In summary, DS leads to lower speed indexes for web pages.}

\noindent {Q: Does DS makes Web page loading slower?}

Figure~\ref{fig:perf}(ii) and Figure~\ref{fig:perf}(ii) show the timing metrics while accessing over Wi-Fi and Cellular 4G network. For Moto X i.e., Motorola's high end smart phone, the median DS HTML loads in 945 ms over Wi-Fi and 1.07 ms over 4G. The median time when flip to real web page occurs is 5250 ms over Wi-Fi and 6717 ms over 4G. Conservatively, comparing this to the median of direct web page load time ( i.e., without proxy) of 5093 ms over Wi-Fi and 5145 ms over 4G, the overhead is only 3\%-5\% because we pay some price by loading the DS HTML first via DS proxy. However, comparing this to the proxy, this overhead is further reduced to 1.7\%-2\%. The perceived web page load time is reduced by a factor of 5x over Wi-Fi to 6x over 4G for the 90th percentile. 

An interesting observation not evident in the figures is that we found no correlation whatsoever between improvements seen in perceived timing metrics and the improvements in speed indexes. This is counter-intuitive at first as the web sites with most complexity must also show similar improvements in speed index but on the contrary, these are usually more popular websites which invest heavily on end user engagement by consciously trying to reduce time to show visually complete page by techniques like keeping landing pages simpler, adaptive prioritized loading the website's visual components etc. {In summary, as opposed to slowing web page loading due to extra processing DS makes it faster.}

\noindent {Q. Does DS perform equally well  with Wi-Fi and 4G LTE?}

Figure~\ref{fig:perf}(i) also shows the difference in improvements in speed index achieved over Wi-Fi vs. 4G where 4G lags Wi-Fi. We attribute this difference in performance to the cellular network as we observed  slower web page load time for same websites over cellular network as shown in Figure~\ref{fig:perf}(iii. Despite the slower cellular network, there is a very small difference in DriveShaft HTML load time. We attribute this to the bursty nature of cellular network and since DS HTML is usually served always within the first burst, it shows very little degradation when compared to Wi-Fi network. {\em In summary, DriveShaft performs equally well for faster Wi-Fi network when compared to comparatively slower 4G network.}

\noindent {Q. Can DS perform for mass produced slower devices?}

The above experiments were carried out using Moto X which we believe as a representative of high end smart phone. Further, to capture the affect of different device capabilities on DriveShaft, we ran experiments using Moto E, i.e., Motorola’s low end phone. We chose the models from the same company launched roughly around the same time frame to minimize affects of hardware differences in our results. To minimize affect of cellular network variability and space in the paper, we only show results using Wi-Fi. Table~\ref{tab:perf2} compares the results measured with on Moto X and Moto E. For Moto E, the median DS HTML loads in 1148 ms despite the fact that some DS snapshots get very large (i.e., 4 MB is not unheard off). The DS load time decreases by 200 ms compared to the Moto E, but the median flip time decreases by 1.4s only. In summary, using a high end smart phone DS pop is 5x faster that time taken to show visually complete web page while directly loading web page. Further, it is important to note that the median, mean and 3rd quantile values for DS pop are all less than two seconds. In order to get worst case performance, we reused the high resolution screen shots taken with the Moto X (more powerful and high screen resolution device) for these experiments. Although, the screen has only a quarter of the pixels and the screen shots could be way smaller, smaller resolution screen shots were not used for these numbers. So, this gives us the worst case performance on Moto E. Using a slower phone the DS pop is 6x faster than direct web page load when compared to Moto X which gains 5x benefit. {\em In summary, less powerful devices benefit more from DriveShaft than powerful devices as DS essentially hides the latency in web loading process which is more likely to be compute bound on less capable devices.}

\noindent {\em Q. Where does DS stand in wake of deployed solutions?}

Figure~\ref{fig:comp}(i) shows comparison SI ratio and web page load achieved with DS time with (i) Flywheel - Google's data compression proxy, (ii) the best possible place to fetch a web page i.e., with on-device cache. Flywheel perform well for its goals to reduce the amount of data but doesn't help latency much. It is surprising to see that DS can potentially be faster in timing than device's cache which highlights that (i) computation is bottleneck as opposed to network in about 30\% of the websites, a result also presented in previous research~\cite{wprof} and (ii) there are elements in web pages with are marked non-cacheable which have to fetched every time regardless. However, in terms of speed index device cache is always outperforms DS which is simply because images, css which are resposible for the visual completeness of web page are more often cacheable than not. We also compare DriveShaft performance with newer browsers available for android which boast their faster performance. Since, pre-render feature is not available in other browsers, chromium telemetry doesn't capture stats from browsers other than chrome and most available web browsers do not expose any developer interface, we had to manually run those experiments. Due to manual nature, we ran those experiments  with a subset of 50 websites and used a proxy which injects java script to collect performance metric which was then stored. We could not compare performance with Puffin browser because to the best of our knowledge, it performs better by executing Javascript in the cloud and hence, it doesn't support JavaScript based metric collection from client device. Amazon Silk browser is only available for Amazon devices making it impossible for us to run experiments. However, our measurements suggest similar improvements as we observed with Google Chrome and more importantly, leads us to believe that DriveShaft improvements are browser agnostic achievin its design goal of creating a solution with minimal foot print.  

\noindent {\em Q. How much additional processing additional processing and network does DS require on client side?}

\noindent {\bf Network:}
DS increases total number of bytes received and transferred from the client because it requires end clients to perform DS post actions but only for clients that are connected to Wi-Fi. From end client perspective, 
\begin{tightitemize}
\item DS leads to increase in the size of web page received due to injection DS client hook which is a constant and negligible compared to size of web pages. Concretely, DS adds a total of additional 93.3. KB = 4063 bytes (DS client hook) + 91,487 bytes (html2canvas Javascript) to an uncompressed web page. This is negligible when compared to the average size of web pages.
\item End client devices need to upload the snapshot, links with their co-ordinates and view port size to the DS server which is directly proportional to the size of total height and the visual complexity of web page rendered on client device. DS client hook captures snapshots only up to twice the size of view port to limit the size of DS Post. The sizes of DS Post components from Moto X are shown in Figure~\ref{fig:comp}(iii).
\end{tightitemize}

\noindent {\bf CPU, Memory and Power:}
DS requires additional processing on the client side added as part of DS Client hook to be executed on client side i.e., (i) capturing the snapshot of web page using final dom object and (ii) traversing dom object to mine the links on the web page which then are used for creation of DS click map. We observed that capturing the snapshot and links is dependent on web page itself and can be variable for different web pages. Similarly, showing DS HTML before the actual web page also incurs overhead in client devices. Figure~\ref{fig:over}(i),(ii) and (iii) compares the CPU load, memory allocation and power consumption respectively on a time line while browsing three representative web sites varying in their complexity in following scenarios : accessing the website (i) directly, (ii)via DS proxy leading to a DS Post and (iii) loading DS HTML before the actual web page. As evident from figure, DS POST leads to (i) additional processing (as seen in CPU load in Figure~\ref{fig:over}(i)) but that takes place after the web page is completely loaded, (ii) a delayed release of memory when compared to direct access (as seen in Figure~\ref{fig:over})(ii)) and (iii) additional power consumption. The difference in resource consumption is only visible in cases of average and high complexity pages suggesting that DS is not good choice for light pages. we observed a average increase in CPU load and power by 3\%-5\% depending on website while memory consumption is increased for a time window between web page load and posting of DS snapshot. However, overheads due to loading DS HTML range from 0.5\%-3\%.

\noindent {Q. How effective is DS desensitization?}

Figure~\ref{fig:desen} shows desentization steps of a real web page. This shows the effectiveness of a simple solution by highlighting removal of potentially sensitive information in this case - Ads, greetings. Desenitization being a subjective matter, the empirical results observed from the snapshot database on DS server after continuously accessing web pages via DS proxy in loop are shown in Table~\ref{fig:comp}(ii). The results are in agreement with our observation of change in websites. However, important to note is the observed false negatives i.e., wrongly unchanged pixels are considerably less false positives i.e., wrongly masked, which simply means that it is less likely to reveal sensitive information than to mask the insensitive information. Also, important is that we observed only 9.7\% median and 11.7\% 90th percentile of web pages which caused Desensitizer to discard the user posted snapshots. However, it is relatively easy for users to submit fake renderings of snapshot i.e., an attacker can submit multiple times - it would be allow them to cause clients to render an arbitrary image. Alternatively, they could upload a single time and cause the desensitization process to discard the entire image. A simple solution is to use heuristics per web site on how much it changes and discard snapshots if the change is above a threshold. Detailed exploration of the security aspect of DS system remains part of our future work.

%% file: relwk.tex
\section{Technology concerns for DriveShaft}
\label{sec:disc}
In this section, we discuss technology specific concerns and their mitigation strategy.

\noindent {\bf Pre-render browser feature}
DriveShaft relies on chrome's capability to load the requested web page in background. At present, this feature is supported in Google Chrome only. This simply means that DriveShaft can only be used in Chrome until apre-render or similar feature is added in other browsers. However, recent developer activity in Mozilla Firefox~\cite{mozillapre-render} leads us to believe that it is a reasonable assumption.

\noindent {\bf Secure Content - HTTPS}
DriveShaft requires run time injection of Java script code to retrieve the web page snapshot. In doing so, the inherent assumption is that the DS proxy is able to intercept and modify the web page which is not possible due to end to end encryption of payload in secure HTTPS connections. This leads to a impression that DriveShaft is rendered useless in HTTPS. However, for popular web sites which utilize Content Delivery Networks (CDNs), the secure connection is terminated at the CDNs nodes hence making it possible for DriveShaft to seamlessly work.

\section{Related Work}
\label{sec:related}
Previous work to accelerate mobile web performance has mainly focused on optimizing steps in web page loading process reducing DNS lookup time~\cite{dnsopt1,dnsopt2}, re-using persistent connections, web object pre-fetching~\cite{prefetching1,prefetching2}, remote java script execution~\cite{offloading}, rendering\cite{grigorik2013high}, Pre-calculating dependencies~\cite{wprof} to reduce idle time, etc. New protocols such as SPDY~\cite{Wang:2014:SS:2616448.2616484}, HTTP 2.0, QUIC try to handle it at transport layer including other flow control mechanisms~\cite{DRZ,HTQ,AFC} to address the limitations of low network throughput. Instead,\cite{A3} argues that such benefit gets easily nullified if application behavior is not known or CPU resources are not available on mobile devices. DriveShaft is a completely orthogonal approach from all these approaches and tried to topple the problem on its head. 

Internet Explorer 11 introduces a new feature, Back Navigation Cache~\cite{IE11} to increase loading speed for the web pages that the browser has loaded once before. DriveShaft, however, takes it to the next level by leveraging CDN infrastructure to use fully rendered web pages on one client device to service others. DS relies on co-operation between client, CDNs and web site developers to benefit other clients. \cite{NETDEDUP} also presents such cooperation between mobile devices and the network infrastructure to carry out network de-duplication more efficiently. \cite{Bartlett95experiencewith}, \cite{Bartlett:1994:WWW:1439278.1440056} were the earliest solutions similar to DriveShaft in that the system send the images to an Apple Newton to enable users to browse the Web wirelessly from an early mobile device. 

%% file: conclusion.tex
\section{Conclusion}
\label{sec:conc}
We presented DriveShaft - an system that accomplishes seemingly impossible task of decoupling the web page visual response time from its structural complexity. We outlined the important design considerations for such a system. Experimental evaluations show that DriveShaft reduces speed index for websites by a factor of up to 5x while giving a perception of 5x-6x faster page loading. These improvements are visible across different devices, network and browsers without requiring any change in websites, browsers or action from end user and with reasonable overheads. Generally, any networked system generating a user perceivable output from complex steps could benefit from an approach similar to DriveShaft. We believe that going forward such decoupled approaches will become important to reduce the burden on developers to design and/or optimize each and every aspect of a complex networked system.